\documentclass[twocolumn,aps,pre,showpacs]{revtex4}
\usepackage[T1]{fontenc}
\usepackage[latin1]{inputenc}
\usepackage{graphics}

\makeatletter

\providecommand{\LyX}{L\kern-.1667em\lower.25em\hbox{Y}\kern-.125emX\@}

\usepackage[T1]{fontenc}
\usepackage[latin1]{inputenc}


\makeatother

\begin{document}

\title{Locating Pollicott-Ruelle resonances in chaotic dynamical systems: \\
A class of numerical schemes. }

\author{R. Florido}

\affiliation{Departamento de Física Fundamental II. Universidad de La Laguna, 38205 Tenerife,
Spain.\\
Departamento de Física. Universidad de Las Palmas de Gran Canaria, 35017 Gran
Canaria, Spain.}

\author{J.M. Martín-González}

\affiliation{Departamento de Física. Universidad de Las Palmas de Gran Canaria, 35017 Gran
Canaria, Spain.}

\author{J.M. Gomez Llorente}

\email{jmgomez@ull.es}

\affiliation{Departamento de Física Fundamental II. Universidad de La Laguna, 38205 Tenerife,
Spain.}


\begin{abstract}
A class of numerical methods to determine Pollicott-Ruelle resonances in chaotic
dynamical systems is proposed. This is achieved by relating some existing procedures
which make use of Padé approximants and interpolating exponentials to both the
memory function techniques used in the theory of relaxation and the filter diagonalization
method used in the harmonic inversion of time correlation functions. This relationship
leads to a theoretical framework in which all these methods become equivalent
and which allows for new and improved numerical schemes.  
\end{abstract}
\pacs{05.45.Ac, 05.45.Pt, 05.45.Mt}
\maketitle

\section{Introduction}

\label{sec1}The Frobenius-Perron (FP) operator plays a central role in the
statistical analysis of chaotic dynamical systems by ruling the time evolution
of distribution functions (probability densities) in phase space. In flows,
it leads to a continuity differential equation known as the Liouville equation.
The FP propagator admits a Hilbert space representation as a unitary operator,
whose spectrum must therefore belong to the unit circle. The structure of the
corresponding spectral decomposition and the nature of the phase space dynamics
are intimately connected. Besides, this same structure determines the behavior
of time correlation functions and their frequency spectral densities. For instance,
the FP operator for the motion on an \( n \)-torus in an integrable Hamiltonian
flow has a purely discrete spectrum with eigenvalues given by \( e^{i\, {\bf k\Omega }\, t} \),
where \( \mathbf{k} \) is any \( n \)-vector of integer numbers and \( {\bf \Omega } \)
is the \( n \)-vector of the torus fundamental frequencies; time correlation
functions are in this case quasiperiodic and the corresponding spectral densities
present delta-function singularities at certain frequency values from the set
\( \omega ={\bf k\Omega } \). On the other hand, in a mixing dynamical system,
if we exclude the eigenvalue 1 (which is simply degenerate and corresponds to
the invariant measure), the spectrum is continuous; time-correlation functions
must therefore decay. In many discrete maps and continuous flows with chaotic
dynamics this decay is exponential; however, even in this case, the correlation
functions may present strong oscillatory modulations; these lead to bumps in
the corresponding frequency spectral densities which have been interpreted as
resonances, i.e. poles in those functions after their analytical continuation
into the complex frequency plane. From the original theoretical work carried
out on this problem by Pollicott \cite{poll} and Ruelle \cite{ruelle} for
a class (Axiom-A) of chaotic dynamical systems these singularities are generally
known as Pollicott-Ruelle (PR) resonances. If, as happen in this class of systems,
the location of the singularities in the complex frequency plane is an intrinsic
property of the dynamical system, i.e. independent of the observable monitored,
then these resonance poles may be considered to be in correspondence with generalized
eigenvalues of the FP operator \cite{gaspard}. This interpretation requires
extensions of this operator which hold for either positive or negative times,
and which make \( U \) a non unitary propagator whose spectral decomposition
may be written in terms of the biorthonormal basis set of its left and right
eigenstates. A physical way of defining the generalized eigenvalue problem is
as an usual Hilbert space spectral analysis of the coarse-grained dynamics in
the limit of zero coarse graining. This particular way establishes a correspondence
between the coarse-grained Liouvillian dynamics in a chaotic flow and the classical
diffusion equation for disordered systems \cite{alt}. Such a correspondence
is the origin of a new approach to carry out the statistical analysis of the
eigenvalue spectrum of a quantum system whose classical limit is chaotic, which
is an extension of the field theoretical methods used for quantum disordered
systems \cite{field}. In this way the role played by the FP operator and its
resonances is important not only in the statistical analysis of classical chaotic
systems but also in that of their quantum counterparts \cite{gomez}. 

The determination of the leading PR resonances is a necessary step in those
statistical analyses. In general, in order to extract them one has to resort
to numerical approximate schemes. However, the numerical approaches followed
so far are few and not always well founded both mathematically and physically.
The theoretically most complete treatments are also the most involved numerically.
The methods based on the diagonalization of a coarse-grained FP operator in
the Hilbert space of square integrable functions belong to this class. In the
limit of zero coarse-graining and a complete basis set one would obtain the
resonances from the eigenvalues. In order to obtain this limit accurately, repeated
diagonalization of a very large matrix (e.g. 8100\( \times  \)8100 for the
simple standard map \cite{blum}) at different values of the coarse-graining
parameter are required; in some cases this parameter is already set by the matrix
dimension \cite{weber}. A different approach also in this class, uses a periodic
orbit representation of the Fredholm determinant of the FP operator, which can
be expressed in terms of the traces of this operator at different times; the
location of the resonances are then obtained from the zeros of the corresponding
Zeta function \cite{gaspard,alonso,cycle}. The scheme requires the knowledge
of a large number of periodic orbits, which can become a very difficult task
in many real systems.

We already know that the decay behavior of time correlation function between
two observables and the analyticity properties of the corresponding frequency
spectral densities are intimately related to the generalized eigenvalues of
the FP operator. This is the origin of a category of methods in which one actually
carries out an analytical continuation of the spectral density into the complex
frequency plane in order to find the poles associated to the resonances contributing
simultaneously to the two observables chosen. Only a few examples that take
this approach can be found in the literature. It was first followed by Isola
for the Hénon map \cite{isola} and later by Baladi et al. \cite{baladi} for
intermittent systems. As will be seen in this paper, the variational method
proposed by Blum and Agam \cite{blum} also belongs to this class. This scheme,
unlike the class of methods discussed in the previous paragraph, requires very
small computational efforts. For instance, Isola \cite{isola} finds the leading
resonances of the Henon map from a {[}\( L,M \){]} Padé approximant to the
spectral density with \( 16\leq L+M\leq 26 \); similarly, Blum and Agam \cite{blum}
obtain the leading resonances of the cat and standard maps as the eigenvalues
of a \( 4\times 4 \) matrix. However, the mathematical and physical foundations
of this approach are not as deep as in the previous methods. This is seemingly
due to the lack of a general theoretical framework from which one can derive
not only these approximate numerical schemes, but also error and convergence
criteria. Finding such a framework is the main goal of this work. It was achieved
by relating these methods to the memory function techniques used in the general
theory of relaxation \cite{memory}, or to the filter diagonalization approach
followed in the location of quantum resonances \cite{neuhauser,taylor}. A new
class of numerical methods to determine  PR resonances comes out from this theoretical
framework. 

The paper is organized in the following way. In Section \ref{sec2} we will
present the results of the memory function method which are more relevant to
our study. These techniques are usually applied to Hamiltonian, Liouvillian,
and general relaxation operators \cite{memory}. In deterministic dynamical
systems, this would imply dealing with the Liouville operator. However in this
case the memory function scheme does not perform the analytic continuation required
in the determination of the PR resonances unless a coarse grained version of
that operator is used. If instead one chooses to analyze the FP operator this
coarse graining is not needed; but in this case we will require a particular
implementation of the memory function methods to deal with propagators. This
is all performed in Section \ref{sec2}. We will show, for instance, how Padé
approximants appear naturally in this scheme. Filter diagonalization \cite{neuhauser,taylor}
is another approach which is specifically adequate to our problem. This is a
particular formulation of the harmonic inversion problem as an eigenvalue equation.
In section \ref{sec3} we give an account of this method and show its total
equivalence with the memory function scheme. The results of these two sections
set up a theoretical framework for a new class of numerical schemes to determine
PR resonances in chaotic systems. Besides, we prove in Section \ref{sec4} that
all these procedures are equivalent linear formulations of the non-linear numerical
problem known as interpolation by exponentials \cite{baladi,henrici}. This
connection provides new tools to perform a better analysis of issues such as
convergence and numerical stability, which are relevant to establish the reliability
of the numerical approach presented here to locate PR resonances. From such
analysis improved schemes may be designed, as the one that we propose in this
section based on a least squares fit method. The application of these methods
is illustrated in simple dynamical systems in Section \ref{sec5}. Finally,
a summary is presented in Section \ref{sec6}.

\section{The memory function technique}

\label{sec2}As mentioned in the introduction, the direct application of this
technique to deterministic dynamical systems would imply dealing with a coarse
grained Liouville operator. If instead one chooses to analyze the Frobenius-Perron
operator the coarse graining is not generally needed. In this section we will
proceed to perform the required particular implementation of the memory function
methods to deal with this propagator. 

First of all, we have to make the following choice for the scalar product between
observables, 
\begin{equation}
\label{eq1}
\langle f\mid g\rangle =\int f^{*}gd\mu ,
\end{equation}
where \( \mu  \) is the natural invariant measure for the dynamics. The time
evolution of the system is ruled by its Frobenius-Perron operator \( U \).
In the case of maps this corresponds to one iteration step; for flows, \( U \)
may correspond either to the Poincaré map or to a given finite-time step. We
then have
\begin{equation}
\label{eq2}
|f(n+1)\rangle =U|f(n)\rangle .
\end{equation}
Making use of this notation the autocorrelation function \( C(n) \) for an
observable \( f \) reads 
\begin{equation}
\label{eq3}
C(n)=\langle f(0)\mid f(n)\rangle =\langle f(0)\mid U^{n}\mid f(0)\rangle .
\end{equation}

We define a resolvent \( G(\omega ) \) of \( U \) by writing

\begin{eqnarray}
G(\omega )|f(0)\rangle  & = & \sum ^{\infty }_{n=0}e^{i\omega n}|f(n)\rangle \nonumber \\
 & = & \sum ^{\infty }_{n=0}\left( e^{i\omega }U\right) ^{n}|f(0)\rangle ,\qquad \textrm{Im}\, \omega >0.\label{eq4} 
\end{eqnarray}
Therefore, 
\begin{equation}
\label{eq5}
G(\omega )=\frac{1}{1-e^{i\omega }U}\, .
\end{equation}

In the memory function formalism \cite{memory} one chooses a particular state
\( |f\rangle \equiv |f(0)\rangle  \) and considers the diagonal matrix element
\( G_{ff}(\omega )=\langle f\mid G(\omega )\mid f\rangle  \). This is an analytic
function in the upper half plane of the complex frequency \( \omega  \). Its
singularities, generally in the form of single poles at \( \omega =\omega _{i} \),
can appear in the lower half plane. The eigenvalues of \( U \) in term of these
poles are therefore \( e^{-i\omega _{i}} \). Recursive projective techniques,
first developed by Zwanzig \cite{zwanzig} and Mori \cite{mori}, are then implemented
to express \( G_{ff}(\omega ) \) as a continued fraction. The process starts
out by defining two complementary generalized projection operators \( P_{0}=\frac{|f_{0}\rangle \langle \widetilde{f}_{0}|}{\langle \widetilde{f}_{0}\mid f_{0}\rangle } \)
and \( Q_{0}=1-P_{0} \), with the identifications \( |f_{0}\rangle \equiv |f\rangle  \)
and \( \langle \widetilde{f}_{0}|\equiv \langle f| \). This partition is then
used to derive a Dyson type equation 
\begin{equation}
\label{eq6}
P_{0}G_{0}P_{0}=\frac{P_{0}}{P_{0}G_{0}^{-1}P_{0}-P_{0}G^{-1}_{0}Q_{0}G_{1}Q_{0}G^{-1}_{0}P_{0}}\, ,
\end{equation}
with \( G_{0}\equiv G(\omega ) \) and \( G_{1}=\left( Q_{0}G_{0}^{-1}Q_{0}\right) ^{-1} \);
from this equation one readily obtains 
\begin{equation}
\label{eq7}
\langle \widetilde{f}_{0}\mid G_{0}\mid f_{0}\rangle =\frac{\langle \widetilde{f}_{0}\mid f_{0}\rangle ^{2}}{\langle \widetilde{f}_{0}\mid G^{-1}_{0}\mid f_{0}\rangle -e^{2i\omega }\langle \widetilde{f}_{1}\mid G_{1}\mid f_{1}\rangle }\, ,
\end{equation}
 where \( |f_{1}\rangle =Q_{0}U|f_{0}\rangle  \) and \( \langle \widetilde{f}_{1}|=\langle \widetilde{f_{0}}|UQ_{0} \).
The procedure can now be repeated for the diagonal element \( \langle \widetilde{f}_{1}\mid G_{1}\mid f_{1}\rangle  \)
appearing in Eq. (\ref{eq7}); from successive iterations a hierarchy of left
\( \langle \widetilde{f}_{n}| \) and right \( |f_{n}\rangle  \) states and
corresponding projection operators \( P_{n}=\frac{|f_{n}\rangle \langle \widetilde{f}_{n}|}{\langle \widetilde{f}_{n}\mid f_{n}\rangle } \)
is constructed according to the recursion scheme 
\begin{eqnarray}
|f_{n}\rangle  & = & U|f_{n-1}\rangle -a_{n-1}|f_{n-1}\rangle -b^{2}_{n-1}|f_{n-2}\rangle ,\nonumber \\
\langle \widetilde{f}_{n}| & = & \langle \widetilde{f}_{n-1}|U-a_{n-1}\langle \widetilde{f}_{n-1}|-b^{2}_{n-1}\langle \widetilde{f}_{n-2}|,\label{eq8} 
\end{eqnarray}
 with \( \langle \widetilde{f}_{n}|=|f_{n}\rangle =0 \) for negative \( n \),
and where

\begin{equation}
\label{eq9}
a_{n}=\frac{\langle \widetilde{f}_{n}\mid U\mid f_{n}\rangle }{\langle \widetilde{f}_{n}\mid f_{n}\rangle };\quad b^{2}_{n}=\frac{\langle \widetilde{f}_{n}\mid f_{n}\rangle }{\langle \widetilde{f}_{n-1}\mid f_{n-1}\rangle };\quad b_{0}=1.
\end{equation}
 From these results one can easily prove that the states \( \langle \widetilde{f}_{n}| \)
and \( |f_{n}\rangle  \) form a biorthogonal set, i.e \( \langle \widetilde{f}_{n}\mid f_{m}\rangle =0 \)
for \( n\neq m \) and that the values of the elements \( a_{n} \) and \( b_{n} \)
can all be obtained from those of the autocorrelation function \( C(n) \).
The operators \( P_{n} \) and \( Q_{n}=1-P_{n} \) are not self-adjoint in
general. However, the previous analysis only requires them to satisfy \( P^{2}_{n}=P_{n} \)
and \( Q^{2}_{n}=Q_{n} \).

The recursive procedure leads to the following expression for the diagonal matrix
element of the resolvent as a continued J-fraction
\begin{eqnarray}
 &  & R(z)\equiv G_{ff}(\omega )\nonumber \\
 & = & \frac{b^{2}_{0}}{1-a_{0}z\: -\quad }\, \frac{b^{2}_{1}z^{2}}{1-a_{1}z\: -\quad }\, \frac{b^{2}_{2}z^{2}}{1-a_{2}z\: -\quad }\cdots \label{eq10} 
\end{eqnarray}
 where \( z=e^{i\omega } \).

From equations (\ref{eq3}), (\ref{eq4}) we find readily the expression relating
this diagonal element \( R(z) \) to the corresponding autocorrelation function
\( C(n) \) 
\begin{equation}
\label{eq11}
R(z)=\sum ^{\infty }_{n=0}C(n)z^{n}.
\end{equation}
The correspondence that we have just found between the power series in Eq. (\ref{eq11})
and the continued fraction in Eq. (\ref{eq10}) is well known in the theory
of continued fractions \cite{fraction}, where a theorem establishes that under
general conditions, such a correspondence is indeed one-to-one. This implies
that the rational function defined by the \( p^{\textrm{th}} \) approximant
\begin{equation}
\label{eq12}
R^{(p)}(z)=\frac{\alpha _{0}+\cdots +\alpha _{p-1}z^{p-1}}{1+\beta _{1}z+\cdots +\beta _{p}z^{p}}
\end{equation}
 to the continued fraction (\ref{eq10}), which is obtained by taking \( a_{n}=b_{n}=0 \)
for \( n\geq p \), has the autocorrelation values \( C(0) \), \( C(1) \),
... \( C(2p-1) \) as its first \( 2p \) Taylor coefficients at \( z=0 \).
As a matter of fact, these are the only values of the autocorrelation function
required in the determination of the elements \( a_{n} \) and \( b_{n} \),
which are needed to obtain from them the \( p^{\textrm{th}} \) approximant
\( R^{(p)}(z) \), as can be deduced from their definition in Eq. (\ref{eq9})
and from the recursive relations in Eq. (\ref{eq8}).

Let us assume for a moment that the autocorrelation function has the following
form as a sum of \( p \) exponentials
\begin{equation}
\label{eq13}
C(n)=\sum _{i=1}^{p}c_{i}z^{n}_{i}.
\end{equation}
In this case the continued fraction expression for \( R(z) \) truncates exactly
at the \( p^{\textrm{th}} \) approximant; in other words \( R(z) \) has an
exact rational representation in the form of Eq. (\ref{eq12}). Then, the poles
of this rational function are the \( p \) complex numbers \( z^{-1}_{i} \),
their corresponding residues being \( -c_{i}z^{-1}_{i} \). The analytic theory
of rational functions shows that the converse is also true \cite{henrici};
namely, if the \( p^{\textrm{th}} \) approximant \( R^{(p)}(z) \) can be computed
from the first \( 2p \) values of an autocorrelation function \( C(n) \),
then each one of these \( 2p \) values satisfy exactly Eq. (\ref{eq13}) with
\( z_{i} \) and \( c_{i} \) given, as before, in terms of the poles and the
residues of \( R^{(p)}(z) \). By fixing the values of the \( 2p \) parameters
\( (z_{i},c_{i}) \) from the first \( 2p \) values of \( C(n) \), we are
indeed performing an interpolation of \( C(n) \) by a sum of \( p \) exponentials
\cite{baladi,henrici}. 

The memory function scheme just presented gives a physical support to the use
of Padé approximants in the representation of the power series for \( R(z) \)
in Eq. (\ref{eq11}). For instance, from the preceding analysis we have 
\begin{eqnarray}
R^{(p)}(z) & = & \frac{\alpha _{0}+\cdots +\alpha _{p-1}z^{p-1}}{1+\beta _{1}z+\cdots +\beta _{p}z^{p}}\nonumber \\
 & = & \sum ^{2p-1}_{n=0}C(n)z^{n}+O(z^{2p}).\label{eq14} 
\end{eqnarray}
Therefore \( R^{(p)}(z) \) gives the {[}\( p-1,p \){]} Padé approximant to
\( R(z) \). In the location of PR resonances, Padé approximants were first
used by Isola in the Hénon map \cite{isola} and later by Baladi et al. \cite{baladi}
in an intermittent system. 

Unlike the method of interpolating exponentials or its equivalent form as a
Padé approximant to \( R(z) \), which can only provide the values of the leading
resonances participating in a given observable, the memory function approach
makes also possible, in principle, the calculation of the corresponding generalized
eigenstates. Indeed, it is not hard to prove (this is the essence of the Lanczos
method \cite{memory}) that in the biorthonormal basis set given by the states
\begin{eqnarray}
|\Phi _{n}\rangle  & = & \langle \widetilde{f}_{n-1}|f_{n-1}\rangle ^{-1/2}|f_{n-1}\rangle ,\nonumber \\
\langle \widetilde{\Phi }_{n}| & = & \langle \widetilde{f}_{n-1}|f_{n-1}\rangle ^{-1/2}\langle \widetilde{f}_{n-1}|,\label{eq15} 
\end{eqnarray}
the evolution operator \( U \) admits the following tridiagonal complex-symmetric
matrix representation 
\begin{equation}
\label{eq16}
U\longleftrightarrow \left( \begin{array}{cccccc}
\; a_{0} & \; b_{1} &  &  &  & \\
\; b_{1} & \; a_{1} & \; b_{2} &  &  & \\
 & \; b_{2} & \; a_{2} &  &  & \\
 &  &  & \ddots  & \; b_{n} & \\
 &  &  & \; b_{n} & \; a_{n} & \\
 &  &  &  &  & \ddots 
\end{array}\right) .
\end{equation}

Again, if \( C(n) \) has the form given in Eq. (\ref{eq13}) as a sum of \( p \)
exponentials, the above matrix truncates exactly into a \( p\times p \) block,
i.e. \( a_{n}=b_{n}=0 \) for \( n\geq p \); from the eigenvalues, \( z_{i} \)
we obtain the location of \( p \) resonances, which coincide exactly with that
found from the poles of the \( p^{\textrm{th}} \) approximant \( R^{(p)}(z) \)
in Eq. (\ref{eq12}). The corresponding right (left) eigenvectors, which are
written as linear combinations of the right (left) states of the biorthonormal
basis set given before, provide a representation of the right (left) generalized
eigenvectors associated to the resonances found. 

In a real dynamical system the correlation function is expected to be dominated
by a few leading resonances, but many others may intervene with small contributions.
Besides, statistical fluctuations are always present in any estimate of \( C(n) \)
from the system orbits. Under such circumstances truncation of the memory function
recursive scheme is required, but one has to choose carefully the order \( p \)
to obtain the location of these resonances with enough accuracy. We will come
back to this important issue later.

\section{Filter diagonalization scheme}

\label{sec3}Filter diagonalization is a particular procedure to solve the problem
of fitting a signal \( C(n) \) to a sum of complex exponentials as in Eq. (\ref{eq13}).
In this approach, this harmonic inversion problem is reformulated as an eigenvalue
problem for an effective evolution operator \( U \). The original formulation
of this scheme \cite{neuhauser} is simplified if, as in our case, the signal
\( C(n) \) is sampled on an equidistant grid \cite{taylor}. Then using our
notation, the matrix representation for \( U \) is realized in the basis set
of Fourier-type states
\begin{equation}
\label{eq17}
|\Psi _{n}\rangle =\sum ^{p-1}_{m=0}e^{im\varphi _{n}}U^{m}|f\rangle ;\quad n=1,2,\ldots ,q,
\end{equation}
where the \( q \) real phases \( \varphi _{n} \) belong to an interval \( \varphi _{min}<\varphi _{n}<\varphi _{max} \)
contained in \( 2\pi >\varphi _{n}\geq 0 \). Then the generalized eigenvalue
problem to solve is 
\begin{equation}
\label{eq18}
\mathbf{UV}=\mathbf{SVZ},
\end{equation}
where the \( \mathbf{U} \) and \( \mathbf{S} \) are complex symmetric matrices,
whose elements are given by 
\begin{eqnarray}
\mathbf{S}_{mn} & = & \left( \Psi _{m},\: \Psi _{n}\right) ,\nonumber \\
\mathbf{U}_{mn} & = & \left( \Psi _{m},\: U\, \Psi _{n}\right) .\label{eq19} 
\end{eqnarray}
 In these equations the symbol \( \left( \bullet ,\bullet \right)  \) defines
a complex symmetric inner product, i.e without complex conjugation. The columns
of the matrix \( \mathbf{V} \) give the eigenvectors 
\begin{equation}
\label{eq20}
|\chi _{n}\rangle =\sum _{m}\mathbf{V}_{mn}|\Psi _{m}\rangle ,
\end{equation}
and \( \mathbf{Z} \) is the diagonal matrix with the complex eigenvalues \( z_{i} \).
Both \( \mathbf{S} \) and \( \mathbf{U} \) usually have rapidly decaying off-diagonal
elements, which can all be calculate from the correlation values \( C(n) \).
Then the eigenvalues \( z_{i} \) lying near a segment \( (e^{i\varphi _{min}},e^{i\varphi _{max}}) \)
of the unit circle in the complex plane may be obtained by solving Eq. (\ref{eq18})
in the basis set of Eq. (\ref{eq17}) restricted to \( q \) different values
of the real phases \( \varphi _{n} \) belonging to the interval \( \varphi _{min}<\varphi _{n}<\varphi _{max} \).
The number \( q \) should be large enough to extract all the leading eigenvalues
in that region of the complex plane. The amplitudes \( c_{i} \) are then calculated
from the eigenvectors as \( c_{i}=\left( f,\, \chi _{i}\right) ^{2} \) \cite{taylor}.

Let us show now the close relationship between memory function and filter diagonalization
schemes. On one hand, note that the use of the complex symmetric inner product
\( \left( \bullet ,\bullet \right)  \) in the latter may be avoided if one
defines, as in the memory function approach, a left functional space expanded
here by the states 
\begin{equation}
\label{eq21}
\langle \widetilde{\Psi }_{n}|=\sum ^{p-1}_{m=0}e^{im\varphi _{n}}\langle f|U^{m};\quad n=1,2,\ldots ,q.
\end{equation}
 If the observable \( f \) is a real function of the phase space variables
(the case of complex \( f \) will be discussed later), then \( \mathbf{S} \)
and \( \mathbf{U} \) matrix elements have the usual form
\begin{eqnarray}
\mathbf{S}_{mn} & = & \langle \widetilde{\Psi }_{m}\mid \Psi _{n}\rangle ,\nonumber \\
\mathbf{U}_{mn} & = & \langle \widetilde{\Psi }_{m}\mid U\mid \Psi _{n}\rangle .\label{eq22} 
\end{eqnarray}
 On the other hand, we have shown in the previous section that the memory function
scheme can be cast as an eigenvalue problem of a tridiagonal matrix; this matrix
is a representation of the evolution operator \( U \) in a biorthonormal basis
set whose states are obtained through a recursive procedure. From equations
(\ref{eq8}) and (\ref{eq15}) we note that for any order \( p \) these states
can be written as linear superpositions
\begin{eqnarray}
|\Phi _{n}\rangle  & = & \sum ^{p}_{m=1}\gamma _{nm}U^{m-1}|f\rangle ,\quad n=1,2,\ldots ,p,\nonumber \\
\langle \widetilde{\Phi }_{n}| & = & \sum ^{p}_{m=1}\gamma _{nm}\langle f|U^{m-1},\quad n=1,2,\ldots ,p,\label{eq23} 
\end{eqnarray}
 with coefficients \( \gamma _{nm} \) obtained recursively. Therefore, if we
abandon the recursive scheme and instead of the states \( \langle \widetilde{\Phi }_{n}| \)
and \( |\Phi _{n}\rangle  \) we take directly 
\begin{eqnarray}
|f(n)\rangle  & = & U^{n}|f\rangle ,\quad n=0,1,\ldots ,p-1,\nonumber \\
\langle \widetilde{f}(n)| & = & \langle f|U^{n},\quad n=0,1,\ldots ,p-1,\label{eq24} 
\end{eqnarray}
to expand the left an right functional spaces we arrive to a generalized eigenvalue
problem equal in form to Eq. (\ref{eq18}) and where the \( p\times p \) \( \mathbf{U} \)
and \( \mathbf{S} \) complex symmetric matrices are now 
\begin{eqnarray}
\mathbf{S}_{mn} & = & \langle f|U^{m-1}U^{n-1}|f\rangle =C(m+n-2),\nonumber \\
\mathbf{U}_{mn} & = & \langle f|U^{m-1}UU^{n-1}|f\rangle =C(m+n-1).\label{eq25} 
\end{eqnarray}
 The columns and rows of the matrix \( \mathbf{V} \) give respectively the
right and left eigenvectors 
\begin{eqnarray}
|\chi _{n}\rangle  & = & \sum ^{p}_{m=1}\mathbf{V}_{mn}|f(m-1)\rangle ,\nonumber \\
\langle \widetilde{\chi }_{n}| & = & \sum ^{p}_{m=1}\mathbf{V}_{mn}\langle \widetilde{f}(m-1)|.\label{eq26} 
\end{eqnarray}
 The variational approach followed by Blum and Agam \cite{blum} to locate the
leading PR resonances in the standard and perturbed cat maps is a \( 4\times 4 \)
implementation of this eigenvalue problem. 

Notice that the two basis sets defined in equations (\ref{eq21}) and (\ref{eq24})
are particular choices of the general form given in Eq. (\ref{eq23}), i.e \( \gamma _{mn}=\delta _{mn} \),
and \( \gamma _{mn}=e^{im\varphi _{n}} \), respectively. Therefore, both eigenvalue
problems can be made equivalent if there exists an invertible linear transformations
between states \( |\Phi _{n}\rangle  \) and \( |\Psi _{n}\rangle  \). This
occurs if \( q \), the number of \( \varphi _{n} \) phases in Eq. (\ref{eq17}),
is chosen \( q=p \), and the values \( e^{i\varphi _{n}} \) sample conveniently
the whole unit circle. The number of correlation points \( C(n) \) required
to set up the corresponding \( p\times p \) eigenvalue problems is always \( 2p \). 

The methods described in this and previous sections can be easily generalized
to cross-correlation signals \( C_{gf}(n)=\langle g(0)\mid f(n)\rangle  \).
In this case, the states expanding the right and left eigenstates are respectively
\begin{eqnarray}
|f(n)\rangle  & = & U^{n}|f(0)\rangle ,\nonumber \\
\langle \widetilde{g}(n)| & = & \langle g(0)|U^{n},\label{eq27} 
\end{eqnarray}
which lead to the same kind of generalized eigenvalue problem. As a matter of
fact, the filter diagonalization analysis of a complex observable \( f \) corresponds
in our notation to a cross-correlation function \( C_{gf}(n) \) with \( g=f^{*} \).
If a full multichannel cross-correlation matrix \( C_{i,j}(n) \) is available
for a set of observables \( f_{i} \) the basis set expanding the right and
left spaces may be obtained as direct sums of those corresponding to each observable
\( f_{i} \) (see also \cite{neuhauser,cross}). 

With the results in this section we conclude in our goal of finding a unified
physical framework for certain schemes followed to find PR resonances which
make use of techniques such as Padé approximants and diagonalization of small
dynamically adapted eigenvalue problems. This framework clearly allows for a
better analysis of issues such as convergence and numerical stability, which
are relevant to establish the reliability of this numerical approach in locating
PR resonances. These and other related aspects will be discussed in the following
sections.

\section{Interpolation by exponentials as an eigenvalue problem}

\label{sec4}In Section \ref{sec2} we have already anticipated that when we
truncate the hierarchy of equations appearing in the memory function scheme
we are in fact carrying out an interpolation of the correlation function by
a finite sum of exponentials \cite{henrici}. In this section we will proceed
to give a general proof of such statement, namely we will show that the method
of interpolating exponentials can be recast as a generalized eigenvalue problem
equivalent to that of the preceding section. 

Suppose we want to solve the harmonic inversion problem by interpolating a given
set of correlation data \( C(n) \) by the sum of \( p \) complex exponentials
\( \sum _{i=1}^{p}c_{i}z_{i}^{n} \), where the unknown parameters are the \( p \)
amplitudes \( c_{i} \) and the \( p \) numbers \( z_{i} \). If we use the
\( 2p \) first values of \( C(n) \), those parameters can be obtained, in
principle, from the solution of the set on non-linear equation
\begin{equation}
\label{eq28}
C(n)=\sum _{i=1}^{p}c_{i}z^{n}_{i},\quad n=0,1,\ldots ,2p-1.
\end{equation}
 Using these \( 2p \) correlation values we can obtain the \( p\times p \)
matrices \( \mathbf{S} \) and \( \mathbf{U} \) defined in Eq. (\ref{eq25}).
From this equation and the set (\ref{eq28}) we derive the following expressions
to be satisfied by these matrices 
\begin{eqnarray}
\mathbf{S} & = & \mathbf{W}^{\top }\mathbf{W},\nonumber \\
\mathbf{U} & = & \mathbf{W}^{\top }\mathbf{ZW},\label{eq29} 
\end{eqnarray}
 where \( \mathbf{W} \), and \textbf{\( \mathbf{Z} \)} have matrix elements,
\( \mathbf{W}_{ij}=c^{1/2}_{j}z^{j-1}_{i} \), \( \mathbf{Z}_{ij}=z_{i}\delta _{ij} \),
and \( \mathbf{W}^{\top } \) denotes the transpose matrix. We finally obtain
from Eq. (\ref{eq29}). 
\begin{equation}
\label{eq30}
\mathbf{UW}^{-1}=\mathbf{SW}^{-1}\mathbf{Z},
\end{equation}
which is identical to the eigenvalue problem of the preceding section. Since
\( \mathbf{W} \) and \textbf{\( \mathbf{Z} \)} on one hand, and \( \mathbf{S} \)
and \( \mathbf{U} \) on the other depend on the same number of parameters (\( 2p \)),
the eigenvalue problem in Eq. (\ref{eq30}) is equivalent to the system in Eq.
(\ref{eq28}). 

A similar eigenvalue problem can be set up such that it is equivalent to the
interpolation of the elements of a multichannel cross-correlation matrix \( C_{ij}(n)=\langle f_{i}(0)\mid f_{j}(n)\rangle  \)
by a sum of the same \( p \) exponentials, i.e. 
\begin{eqnarray}
C_{ij}(n)=\sum _{k=1}^{p}c_{ij,k}z^{n}_{k}, &  & n=0,1,\ldots ,2p_{ij}-1,\nonumber \\
 &  & i,j=1,2,...,q,\label{eq30b} 
\end{eqnarray}
 with the restriction \( c_{ij,k}=\widetilde{a}_{ik}a_{jk} \). 

The existence of solution to the problem in Eq. (\ref{eq30}) may be guaranteed
by requiring that both \( \mathbf{S} \), and \textbf{\( \mathbf{U} \)} be
nonsingular matrices. A singular \( \mathbf{S} \) matrix would reveal linear
dependences in the basis set, and this will always occur when the order \( p \),
i.e. half the number of correlation data used, is larger than the true number
of exponentials required to expand \( C(n) \). A singular \textbf{\( \mathbf{U} \)}
would imply a zero eigenvalue \( z_{i} \), which is usually linked to a singular
\( \mathbf{S} \) matrix. One can get rid off these linear dependences by reducing
the number of correlation data and thus the number of states until \( \mathbf{S} \)
becomes nonsingular. In the recursive method of the memory function scheme,
if the \( (p+1)\times (p+1) \) \( \mathbf{S} \) matrix becomes singular for
\( p=p_{o} \) then the coefficient \( b_{p_{o}} \) vanishes, thus truncation
occurs naturally. The same results are obtained if one performs a singular value
decomposition of \( \mathbf{S} \) \cite{neuhauser}. In practice, in a real
dynamical system, although the correlation function may be expected to be dominated
by a few leading resonances, many others may intervene with small contributions.
Besides, statistical fluctuations are always present in any estimate of \( C(n) \)
from the system orbits. Under such circumstances there may not exist an order
\( p \) such that \( b_{p} \) vanishes exactly or \( \mathbf{S} \) becomes
exactly singular. In very favorable situations one may identify an optimal order
\( p_{o} \) if a sharp decrease of the magnitude of either \( b_{p} \) or
the \( (p+1)\times (p+1) \) determinant \( \left| \mathbf{S}\right|  \) is
observed at \( p=p_{o} \). We will see later that in chaotic dynamical systems
such a behavior is not generally found, and one should proceed very carefully.
In any case, the optimal order is usually much smaller than the number of correlation
data which can be determined. This fact allows for the implementation of methods
to reduce the statistical noise in \( C(n) \), so that the \( 2p \) values
used to set up the eigenvalue problem become more accurate. The standard procedure
is to perform a least squares fit. 

A general scheme for this procedure can be designed to analyze a full cross-correlation
matrix; however for simplicity, we only present here the case of a single correlation
signal. Our goal is then to perform an optimal fit of \( q \) correlation values
by a sum of \( p \) exponentials, with \( q\gg p \). There are different solutions
to this problem depending on the error function \( \xi  \) to minimize \cite{marple}.
Here we will make the choice
\begin{equation}
\label{eq31}
\xi =\sum ^{q-1}_{n=0}\left| C(n)-\sum _{i=1}^{p}c_{i}z^{n}_{i}\right| ^{2},
\end{equation}
where \( C(n) \) is the known correlation signal.

Let us show now that this non-linear problem can be again reformulated as a
linear one in which an eigenvalue problem must be solved self-consistently.
By minimizing \( \xi  \) with respect to the parameters \( c_{i} \) and \( z_{i} \)
we obtain the system of equations 
\begin{eqnarray}
\sum ^{q-1}_{n=0}z^{n}_{i}\varepsilon _{n} & = & 0,\quad i=1,2,\ldots ,p,\nonumber \\
\sum ^{q-1}_{n=0}nc_{i}z^{n-1}_{i}\varepsilon _{n} & = & 0,\quad i=1,2,\ldots ,p,\label{eq32} 
\end{eqnarray}
 where 
\begin{equation}
\label{eq33}
\varepsilon _{n}=C(n)-\sum _{i=1}^{p}c_{i}z^{n}_{i}
\end{equation}
is the error between the known correlation values and their fit. We now rewrite
the system (\ref{eq32}) in the form
\begin{eqnarray}
\sum ^{2p-1}_{n=0}z^{n}_{i}\varepsilon _{n} & = & u_{i},\quad i=1,2,\ldots ,p,\nonumber \\
\sum _{n=0}^{2p-1}nc_{i}z^{n-1}_{i}\varepsilon _{n} & = & v_{i},\quad i=1,2,\ldots ,p,\label{eq34} 
\end{eqnarray}
where \( u_{i}=\sum _{n=2p}^{q-1}z^{n}_{i}\varepsilon _{n} \) and \( v_{i}=\sum _{n=2p}^{q-1}nc_{i}z^{n-1}_{i}\varepsilon _{n} \). 

The solution to this system can be performed in the following self-consistent
way. Suppose we have an initial guess \( c^{(0)}_{i} \) and \( z^{(0)}_{i} \)for
the \( 2p \) parameter \( c_{i} \) and \( z_{i} \), obtained, for instance,
from the solution of the \( p\times p \) eigenvalue problem set up with the
first \( 2p \) values of the known correlation function \( C(n) \). We use
this guess to calculate
\begin{equation}
\label{eq35}
C_{n}^{(0)}=\sum _{i=1}^{p}c^{(0)}_{i}\left[ z^{(0)}_{i}\right] ^{n},\quad n=0,1,\ldots ,q-1,
\end{equation}
 and the terms \( u_{i} \) and \( v_{i} \) and solve system (\ref{eq34})
to find the first \( 2p \) errors \( \varepsilon _{n}\equiv \varepsilon ^{(0)}_{n} \).
With these values we obtain a corrected form of the correlation function
\begin{eqnarray}
C_{n}^{(j+1)} & = & C^{(j)}_{n}+\eta \left[ C(n)+\varepsilon ^{(j)}_{n}-C_{n}^{(j)}\right] ,\nonumber \\
 &  & \qquad \qquad \qquad \qquad n=0,1,\ldots ,q-1,\label{eq36} 
\end{eqnarray}
where \( j=0 \) in this initial step, and \( \eta  \) is a parameter conveniently
chosen. With the first \( 2p \) values of \( C^{(1)}_{n} \) we reset and solve
the \( p\times p \) eigenvalue problem from which we extract an improved guess
\( c^{(1)}_{i} \) and \( z^{(1)}_{i} \); the procedure is now iterated until
the fix point of the map defined by Eq. (\ref{eq36}) is reached. This will
require a convenient choice of the parameter \( \eta  \). We obtain in this
way the best fit in the sense of the error function in Eq. (\ref{eq31}) and
from the solution of the last eigenvalue problem the best eigenvalues and eigenvectors.
One may start with a small value of the number of exponentials \( p \) and
increase systematically this number fulfilling always the condition \( p\ll q \);
an optimal \( p=p_{o} \) would be the maximum order for which the fixed point
of (\ref{eq36}) can be numerically reached. 

Let us finish this Section with a brief discussion on the physical interpretations
of the left and right eigenfunctions derived numerically from the schemes proposed
in this work. Let us suppose first that the observable chosen \( f \) has a
finite expansion in one or in both of the basis sets corresponding to the left
or right generalized eigenvectors of the FP operator. Then it is straightforward
to see that an exact truncation order \( p=p_{e} \) exists, \( p_{e} \) being
the number of terms in the shortest of the two expansions. Therefore, only the
\( p_{e} \) generalized eigenvectors participating in this shortest expansion
can be obtained form these numerical schemes. More accurate correlation values
will provide better approximations to the exact eigenvectors. The observable
\( f \) being an ordinary function, this situation can occur only if the eigenvectors
of the finite expansion are themselves ordinary functions and not distributions.
We will present an example of this case in the following Section. 

However, in a general chaotic dynamical systems, both the left and right generalized
eigenvectors of the FP operator will be complicated distributions \cite{gaspard}
of the phase space variables. In this case the two expansions of the observable
function \( f \) will not be finite and therefore an exact truncation order
\( p_{e} \) will not exist. As already discussed, we can still obtain a finite
optimal order \( p_{o} \), which will be determined by the accuracy of the
correlation values and by the numerical precision of the computations. The necessary
choice of a finite \( p_{o} \) imposes a time cutoff to the evolution of the
initial state and thus a limit to the phase-space resolution with which the
generalized eigenvectors can be obtained. Therefore, this class of methods provides
in this case smooth representations of the exact eigendistributions. This will
be also illustrated in the next Section.

\section{Numerical examples}

\label{sec5}In this section we will illustrate the use of this new class of
methods by locating the leading PR resonances of some simple chaotic maps. We
first choose the Bernoulli map

\begin{equation}
\label{eq37}
x_{n+1}=2x_{n},\qquad (\textrm{mod}.\: 1),
\end{equation}
for which the resonance values \( z_{i} \) and the generalized left \( \widetilde{\chi }_{i} \)
and right \( \chi _{i} \) eigenstates of the FP operator are known analytically.
These are, respectively (see Ref. \cite{gaspard})
\begin{equation}
\label{eq38}
z_{i}=\frac{1}{2^{i}},\quad i=0,1,2,\ldots 
\end{equation}
and
\begin{eqnarray}
\widetilde{\chi }_{0} & = & \theta (x)\theta (1-x),\nonumber \\
\widetilde{\chi }_{i} & = & (-)^{i-1}\left[ \delta ^{(i-1)}(x-1)-\delta ^{(i-1)}(x)\right] ,\nonumber \\
\chi _{i} & = & \frac{B_{i}(x)}{i!},\label{eq39} 
\end{eqnarray}
 where \( \delta ^{i}(x) \) represents the \( i^{\textrm{th}} \) derivative
of the Dirac delta distribution and \( B_{i}(x) \) are the Bernoulli polynomials.
For the observable \( f(x) \) we take the polynomial function 
\begin{equation}
\label{eq40}
f(x)=x^{3}-\frac{1}{4},
\end{equation}
 and the corresponding normalized autocorrelation function \( C(n)=\frac{\int ^{1}_{0}dxf(x)U^{n}f(x)}{\int ^{1}_{0}dxf^{2}(x)} \)
can be computed analytically from the above spectral decomposition giving 
\begin{equation}
\label{eq41}
C(n)=\frac{14}{15}z_{1}^{n}+\frac{7}{45}z_{2}^{n}-\frac{4}{45}z_{3}^{n}.
\end{equation}
Therefore the three resonances \( z_{i}, \) \( i=1,2,3 \) {[}Eq. (\ref{eq38}){]}
contribute to this correlation function. Of course, in a general practical situation
we will not know the exact correlation function but only an estimate of it,
which is usually obtained from the system orbits; in such case some statistical
error will be always present in \( C(n) \). We simulate these statistical fluctuations
by adding to the exact \( C(n) \) in Eq. (\ref{eq40}) a random noise \( \epsilon  \)
with Gaussian probability distribution 
\begin{equation}
\label{eq42}
P(\epsilon )=\frac{1}{\sqrt{2\pi }\sigma }\exp \left[ -\frac{1}{2}\left( \frac{\epsilon }{\sigma }\right) ^{2}\right] .
\end{equation}
Thus
\begin{equation}
\label{eq43}
C_{r}(n)=C(n)+\epsilon (n),
\end{equation}
and proceed to illustrate the effect of the noise standard deviation \( \sigma  \)
in the implementation of the methods that we have proposed in this work to locate
PR resonances. 

A first parameter to determine is the optimal order \( p_{o} \), i.e the number
of resonances contributing significantly to the correlation function. As mentioned
in Section \ref{sec4}, one may identify an optimal order \( p_{o} \) if a
sharp decrease of the magnitude of either the memory function coefficient \( b_{p} \)
{[}Eq. (\ref{eq9}){]} or the determinant of the \( (p+1)\times (p+1) \) correlation
matrix \( \mathbf{S}(p+1) \) {[}Eq. (\ref{eq25}){]} is observed at \( p=p_{o} \). 
\begin{figure}[!b]
{\par\centering \includegraphics{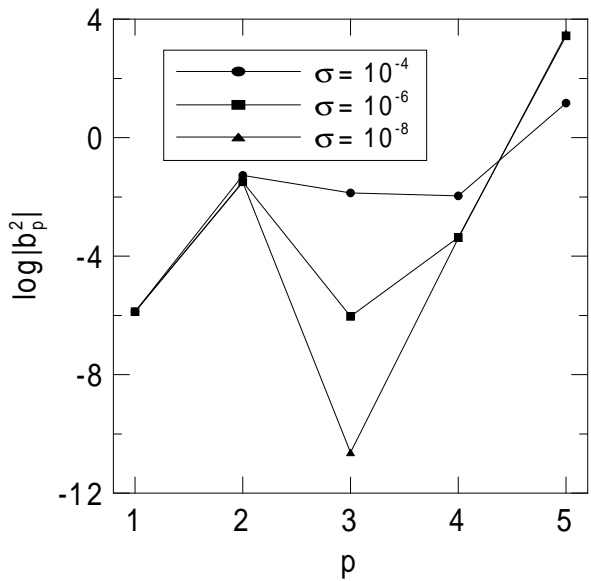} \par}

\caption{\label{fig1} Magnitude of the memory function coefficient \protect\( b^{2}_{p}\protect \)
{[}Eq. (\ref{eq9}){]} as a function of the order \protect\( p\protect \).
These values were obtained for the Bernoulli map from a noisy correlation function
for the observable \protect\( f(x)=x^{3}-\frac{1}{4}\protect \); the parameter
\protect\( \sigma \protect \) gives the standard deviation of the gaussian
random noise. }
\end{figure}
\begin{figure}[!t]
{\par\centering \includegraphics{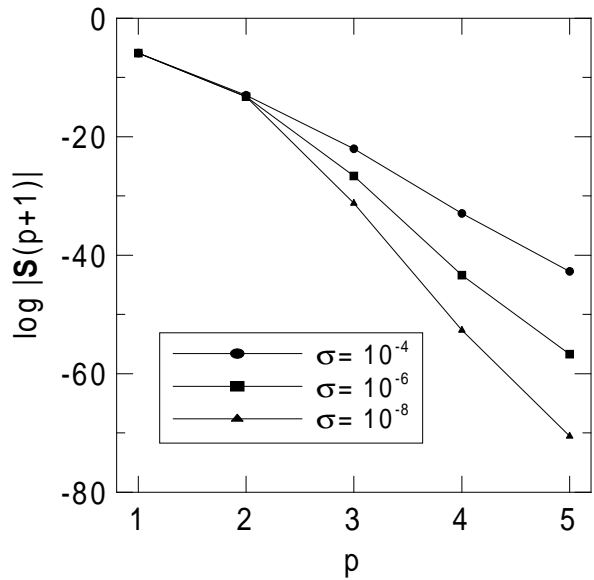} \par}

\caption{\label{fig2} Magnitude of the determinant of the correlation matrix \protect\( \left| \mathbf{S}(p+1)\right| \protect \)
{[}Eq. (\ref{eq25}){]} as a function of the order \protect\( p\protect \).
These values were obtained for the Bernoulli map. Other details are as in Fig.
1.}
\end{figure}
 In figures \ref{fig1} and \ref{fig2} we plot respectively, as a function
of \( p \), the magnitudes of \( b^{2}_{p} \) and \( \left| \mathbf{S}(p+1)\right|  \)
obtained from the noisy correlation function \( C_{r}(n) \) for the Bernoulli
map. We clearly observe in Fig. \ref{fig1} that at the optimal order \( p=p_{o}=3 \)
the coefficient \( b^{2}_{p} \) is very sensitive to \( \sigma  \); the same
sensitivity is observed in \( \left| \mathbf{S}(p+1)\right|  \) for \( p\geq 3 \)
(Fig. \ref{fig2}). Besides, we observe that if the error in the correlation
function is not small enough it may wash out completely the expected sharp decrease
in both \( b^{2}_{p} \) and \( \left| \mathbf{S}(p+1)\right|  \) at \( p=p_{o} \);
the determination of an optimal order would therefore require quite precise
correlation values (\( \sigma <10^{-4} \)). 

If instead one makes use of the self-consistent least-squares method described
in Section \ref{sec4}, \( p_{o} \) would be the order giving the minimum value
of the error function \( \xi  \) defined in Eq. (\ref{eq31}); in general,
\( p_{o} \) coincides with the maximum order \( p \) for which self-consistency
is achieved, i.e. for which the fixed point of Eq. (\ref{eq36}) can be numerically
reached. In the present case, for \( \sigma <10^{-2} \) this occurs always
at the correct value \( p_{o}=3 \), which indicates the ability of the least
squares method to reduce the negative effects of the correlation noise. 

\begin{table} [!b]

\caption{\label{tab1}Resonance locations of the Bernoulli map. They were obtained
from the solution of the eigenvalue problem in Section \ref{sec4}, which was
set up for the autocorrelation function of the observable \( f(x)=x^{3}-\frac{1}{4} \).
The parameter \( \sigma  \) is the standard deviation of the gaussian randon
noise added to the exact correlation values {[}see Eq. (\ref{eq43}){]}, and
\( p \) is the order of the method, i.e. the expected number of resonances
contributing to the correlation function; its correct value is \( p=3 \).} 

\begin{ruledtabular} 

\begin{tabular}{ccccc}\( \sigma  \)&\( p \)&\( z_{1} \)&\( z_{2} \)&\( z_{3} \)\\ 

\hline 

&\( 2 \)& \( 0.4909 \)&\( -03094 \)&\\ 

\( 10^{-4} \)&\( 3 \)&\( 0.4943 \)&\( 0.0881+0.3520i \)&\( 0.0881-0.3520i \)\\ 

&\( 4 \)&\( 0.5370 \)&\( 0.4733 \)&\( -0.1207 \)\\ 

\hline 

&\( 2 \)&\( 0.4909 \)&\( -0.2891 \)&\\ 

\( 10^{-6} \)&\( 3 \)&\( 0.4998 \)&\( 0.2347 \)&\( 0.1379 \)\\ 

&\( 4 \)&\( 0.5001 \)&\( 0.2659 \)&\( 0.0949 \)\\ 

\hline

&\( 2 \)&\( 0.4909 \)&\( -0.2889 \)&\\ 

\( 10^{-8} \)&\( 3 \)&\( 0.5000 \)&\( 0.2499 \)&\( 0.1251 \)\\ 

&\( 4 \)&\( 0.5000 \)&\( 0.2502 \)&\( 0.1251 \)\\

\end{tabular} 

\end{ruledtabular} 

\end{table}

Let us now illustrate the effect of \( \sigma  \) in the resonance locations.
In Table \ref{tab1} we present the resonances derived from the solution of
the eigenvalue problem defined in Section \ref{sec4}, which was set up for
the noisy correlation function \( C_{r}(n) \); different values for \( \sigma  \)
and the order \( p \) were used. We deduce from these results that the leading
resonance \( z_{1} \) is quite robust against noise and non optimal choices
of the order \( p \). The accurate location of other two resonances require,
however, a better selection of \( p \) and a more converged correlation function
estimate; for instance, the \( z_{2} \) and \( z_{3} \) values obtained from
the most noisy correlation (\( \sigma =10^{-4} \)) are completely wrong. 

\begin{table} [!t]

\caption{\label{tab2} Resonance locations of the Bernoulli map. They were obtained
with the self-consistent least-square method described in Section \ref{sec4}.
The optimal order \( p=3 \) given by this method was used. Other details are
as in Table I.} 

\begin{ruledtabular} 

\begin{tabular}{cccc}\( \sigma  \)&\( z_{1} \)&\( z_{2} \)&\( z_{3} \)\\ 

\hline 

\( 10^{-4} \)&\( 0.5003 \)&\( 0.2697 \)&\( 0.0850 \)\\ 

\( 10^{-6} \)&\( 0.5000 \)&\( 0.2505 \)&\( 0.1243 \)\\ 

\( 10^{-8} \)&\( 0.5000 \)&\( 0.2500 \)&\( 0.1250 \)\\ 

\end{tabular} 

\end{ruledtabular} 

\end{table}

The resonance locations obtained from the self-consistent least-squares method
are given in Table \ref{tab2}. As we already know, this method provides as
the optimal order \( p_{o} \) the correct value \( p_{o}=3 \). The method
gives also significantly better resonance eigenvalues even in the case of large
correlation noise (\( \sigma =10^{-4} \)), where the previous approach failed
completely. In conclusion, when statistical errors are present in the estimated
correlation function, of all the methods discussed in this work, the least-squares
scheme provides the most accurate resonance values by reducing the negative
effects of the correlation errors. In general, two factors increase the reliability
of a resonance location obtained in this way: a larger stability and a larger
contribution in the observable chosen. 

The resonance generalized eigenvectors can be also determined from these schemes.
For the Bernoulli map the calculated right eigenvectors are approximations to
the Bernoulli polynomials participating in the observable chosen; a better estimate
of the eigenvalue gives a better approximation to the corresponding Bernoulli
polynomial. However the left eigenvectors provided by these schemes are wrong
representation of the true ones. The reason for such an incorrect result may
lie in the observable chosen and in the very different nature of the exact left
and right generalized eigenvectors: while the right ones are functions, the
left ones are distributions; this asymmetry is a consequence of the non-invertibility
of the map. Therefore, the expansion of our observable (a polynomial) in the
right eigenvector (Bernoulli polynomial) has a finite number of terms while
this number is infinite if such expansion is performed in the left ones. When
these two numbers are different, as in our case, we know from the final discussion
in Section \ref{sec4} that the present schemes will provide a representation
for the states and resonances of the shorter expansion. 

As a second example we will consider the standard map \cite{chirikov}
\begin{eqnarray}
x_{n+1} & = & x_{n}+y_{n},\nonumber \\
y_{n+1} & = & y_{n}+\frac{K}{2\pi }\sin \left( 2\pi x_{n+1}\right) ,\quad (\textrm{mod }1)\label{eq44} 
\end{eqnarray}
This map is area-preserving and invertible. While for \( K=0 \) it is integrable,
for \( K\neq 0 \) presents chaotic phase space regions which become more extensive
as \( K \) increases. The system follows the route to chaos known as overlapping
resonances \cite{chirikov,note}. We take the value \( K=10 \) in our calculations. 

The exact knowledge of the PR resonances is not generally possible for this
system. Blum and Agam \cite{blum} proposed a variational approach to locate
the four leading ones, which as we are already aware, is a \( p=4 \) implementation
of the eigenvalue problem derived in this work. For comparison, we will take
the observable chosen by these authors, i.e. 
\begin{equation}
\label{eq45}
f(x,y)=\exp \left( i2\pi x\right) ,
\end{equation}
whose autocorrelation function may be decomposed as the sum of the autocorrelation
functions for \( \cos 2\pi x \) (\( C_{c} \)) and \( \sin 2\pi x \) (\( C_{s} \))
\begin{equation}
\label{eq46}
C(n)=C_{c}(n)+C_{s}(n).
\end{equation}
 This decomposition is possible because the standard map has an inversion symmetry
with respect to the point \( (x=\frac{1}{2},\; y=\frac{1}{2}) \); therefore,
the generalized eigenvectors are either symmetric or antisymmetric under this
transformation. The symmetric ones will participate in \( C_{c}(n) \), while
the antisymmetric ones will do in \( C_{s}(n) \). 

\begin{table} [!b]

\caption{\label{tab3} Locations of leading resonances in the standard map for
\( K=10 \). The first two lines correspond respectively to \( C_{c}(n) \)
and \( C_{s}(n) \) autocorrelation functions. The resonances were obtained
from these correlation functions using the self-consistent least-squares scheme.
The bottom line gives the four values obtained by Blum and Agam \cite{blum}
from their \( p=4 \) variational approach} 

\begin{ruledtabular} 

\begin{tabular}{lccc}\( f \)&&\( z_{i} \)&\\ 

\hline 

\( \cos (2\pi x) \)&\( 0.672 \)&\( -0.030\pm 0.702i \)&\( 0.332\pm 0.503i \)\\ 

\( \sin (2\pi x) \)&\( -0.715 \)&\( 0.150\pm 0.592i \)&\( -0.119\pm 0.553i \)\\ 

\( e^{i2\pi x} \) \cite{blum}&\( 0.515 \)&\( -0.494 \)&\( -0.003\pm 0.505i \)\\ 

\end{tabular} 

\end{ruledtabular} 

\end{table} 

Both \( C_{c}(n) \) and \( C_{s}(n) \) have been determined here from the
system orbits with an uncertainty \( \sigma <10^{-5} \). We have proceeded
next to implement the self-consistent least-squares method of Section \ref{sec4}
to find the leading PR resonances. As a first result we obtain for the optimal
order the values \( p_{o}=9 \) from \( C_{c}(n) \), and \( p_{o}=8 \) from
\( C_{s}(n) \). The location of the leading resonances is given in Table \ref{tab3}.
Since the two correlation functions do not have common eigenvalues, the total
optimal order corresponding to the initial observable in Eq. (\ref{eq45}) is
\( p_{o}=17 \); this number is much larger than the value \( p=4 \) used by
Blum and Agam. If we take this smaller order to solve the general eigenvalue
problem, the locations that we obtain for the the four resonances coincide exactly
with those found by these authors; they are also included in Table \ref{tab3}.
However, if we increase \( p \) just by one these values change significantly,
which means that they are not properly converged. 

\begin{figure}[!t]
{\par\centering \includegraphics{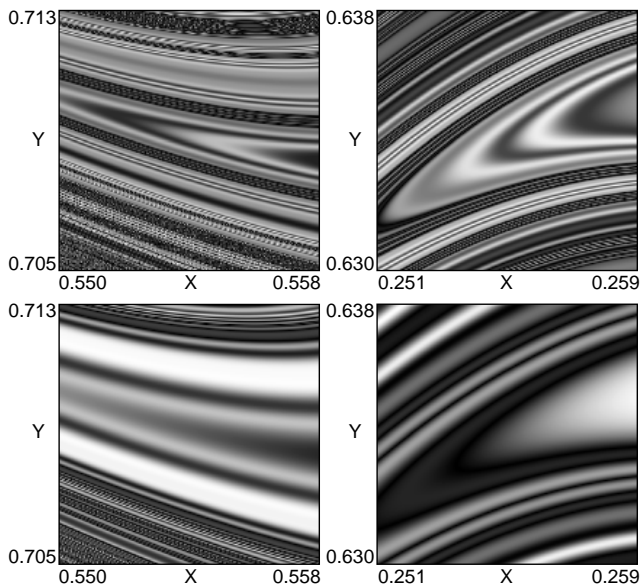} \par}

\caption{\label{fig3}Structure of the right (right panels) and left (left panels) generalized
eigenvectors corresponding to the real resonances \protect\( z_{i}=0.672\protect \)
(top) and \protect\( z_{i}=-0.715\protect \) (bottom) in the standard map (\protect\( K=10\protect \)).
The eigenvectors in the top panels are symmetric under the inversion transformation
and participate in the cosine correlation function \protect\( C_{c}\protect \);
those in the bottom panels are antisymmetric and participate in the sine correlation
function \protect\( C_{s}\protect \). To better illustrate the intricate structure
we have selected small regions of the available phase space. Absolute magnitudes
are represented, with higher values corresponding to brighter regions.}
\end{figure}

The corresponding left and right generalized eigenvectors are expected to be
complicated distributions. In this case, as we have discussed in Section \ref{sec4},
the eigenfunctions derived from our numerical approach are smooth representations
of them. In Fig. \ref{fig3} we present the numerical left and right eigenfunctions
associated with each of the two leading resonances of the standard map; the
first pair is symmetric and the second one antisymmetric under the inversion
transformation. To better illustrate the intricate structure we have selected
small regions of the available phase space. 

\section{Summary}

\label{sec6}Among the schemes followed in the literature to locate the leading
Pollicott-Ruelle resonances of the Perron-Frobenius operator in chaotic dynamical
systems, there are a few examples of remarkable simplicity, which involve either
the use use of Padé approximants to perform the analytical continuation of the
spectral density functions \cite{isola,baladi}, or the diagonalization of small
dynamically adapted eigenvalue problems \cite{blum}. In an attempt to provide
a theoretical support to these methods, we have analyzed their connection with
other numerical schemes used in different contexts. A first category of such
schemes includes the memory function techniques \cite{memory} used in the general
theory of relaxation, and the methods related to this approach such as those
based on the use of continued fractions and Padé approximants. In a second category
we have also considered the methods of the filter diagonalization approach \cite{neuhauser,taylor},
which is a particular formulation of the harmonic inversion of a time signal
as an eigenvalue problem. 

The analysis of these schemes led us to a theoretical framework in which all
them become equivalent formulations of the same problem: the location of the
leading resonances contributing to a give time correlation function. The most
convenient of these formulations is as an eigenvalue problem from which one
can obtain not only the poles, i.e. the resonance locations, but also a smooth
representation of the generalized eigenvectors. Besides, we have proved that
all these procedures are also particular linear formulations of the non-linear
numerical problem known as interpolation by exponentials \cite{baladi,henrici}.
This connection has provided us with new tools to perform a better analysis
of issues such as convergence and numerical stability. From such analysis improved
schemes may be designed, as the one that we have proposed based on a least-squares
fit. 

We have illustrated the use of this class of methods in two chaotic maps: the
Bernoulli map and the standard map. In the first example the nature of the right
eigenvectors, which are known to be the Bernoulli polynomials, allows for finite
expansion of certain observables like the one chosen for illustration. Then
an accurate enough correlation function provides good approximations to the
PR resonances and their corresponding right eigenvectors. The least squares
method improves significantly these results in the case of less accurate correlation
values. 

In the standard map, for which the exact knowledge of the PR resonances is not
generally possible, the more elaborated least squares method provides resonances
values which appear to be significantly more converged than previous results
\cite{blum}. Nice smooth representations of the eigendistributions are also
obtained. These indicate the relevant phase space regions involved in the dynamics.

\begin{acknowledgments}This work has been supported by a grant from ``Ministerio
de Ciencia y Tecnología (Spain)'' under contract No. BFM2001-3343. 

\end{acknowledgments}


\begin{thebibliography}{10}
\bibitem{note}The term resonance here does not mean PR resonance but dynamical resonance between
torus frequencies. \vspace{0.3cm}
\end{thebibliography}
\end{document}